# Fabricating Paper Circuits with Subtractive Processing

Ruhan Yang

University of Colorado Boulder, ruhan.yang@colorado.edu

Krithik Ranjan

University of Colorado Boulder, krithik.ranjan@colorado.edu

Ellen Yi-Luen Do

University of Colorado Boulder, ellen.do@colorado.edu

This paper introduces a new method of paper circuit fabrication that overcomes design barriers and increases flexibility in circuit design. Conventional circuit boards rely on thin traces, which limits the complexity and accuracy when applied to paper circuits. To address this issue, we propose a method that uses large conductive zones in paper circuits and performs subtractive processing during their fabrication. This approach eliminates design barriers and allows for more flexibility in circuit design. We introduce PaperCAD, a software tool that simplifies the design process by converting traditional circuit design to paper circuit design. We demonstrate our technique by creating two paper circuit boards. Our approach has the potential to promote the development of new applications for paper circuits.

## 1 INTRODUCTION

Paper circuits combine electronics and paper-based materials to provide an accessible and customizable platform for a wide range of applications. The ubiquity and flexibility of paper make it an ideal material for creating electronic circuits that can be easily integrated into a variety of designs. In addition to being widely popular with the maker community, researchers such as Qi have also completed a series of studies on paper circuits. Despite the great potential of paper circuits, their current fabrication methods have some key limitations. Traditional paper circuits are designed using a circuit layout based on thin traces, which is the same as that used in rigid printed circuit boards (PCBs). However, the use of thin conductive traces can limit the complexity of the circuit, especially when creating paper circuits by hand. At the same time, high-precision manufacturing methods using specialized equipment or materials are often expensive and not affordable for all. These limit the wider use of paper circuits as a versatile and affordable technology.

To overcome existing limitations and to make paper circuits more accessible and customizable, we present a new fabrication technique that uses large conductive zones and subtractive processing to create complex and precise paper circuits. In addition, we introduce PaperCAD (Paper Circuit Art Design), a software tool that converts traditional circuit design to paper circuit design, making the design process easier. Our goal is to bring electronics to a wider audience, enabling people with different skill levels to design and fabricate their own electronic circuits with paper. By proposing new fabrication methods and developing the new software tool, our goal is to create a low-cost, accessible, and environmentally friendly platform for creating customized circuit boards. In this position paper, we present the design and use of this technique and discuss the insights we have gained from it.

## 2 RELATED WORKS

Many researchers have explored the combination and integration of paper and electronics. Researchers have proposed different ways to deposit conductive inks on paper, such as hand drawing [Buechley, 2009] and printing circuits on paper [Saul 2010]. Others have explored the various functions that can be achieved by adhering conductive tape on paper [Qi, 2010]. All of these studies have shown us the feasibility of paper-based circuits; however, one of the constraints they face is that the precision and complexity of these circuits are usually determined by the tools and materials.

In order to simplify circuit design and fabrication, some researchers have investigated the use of pre-processed materials, such as cut-out patterned plastic sheet material to help with drawing [Li, 2016]. Specialized printing tools can greatly improve the precision and complexity of these drawn circuits [Saul 2010]. Haider [2013] created software to connect to a cutting plotter machine for making copper prints on paper. Recent work [Qi, 2021] proposes the use of vinyl cutters to make circuits based on painting, which is a more accurate way to make circuits using prefabricated conductive materials such as copper foil tape.

With all the paper circuit fabrication methods proposed so far, a very significant common feature is that they all use additive processing for fabrication. However, if we compare the fabrication of paper circuits to the production process of printed circuit boards (PCBs), we can see that an alternative manufacturing technique is subtractive processing [Coombs, 2008], which is not yet widely discussed in paper circuit fabrication. Subtractive processing offers several advantages over traditional additive processing for manufacturing PCBs. It allows for greater accuracy and design flexibility, as well as improved adaptability to different materials. In addition, since it does not require expensive equipment or materials, it is a cost-effective method for small-scale production, making it a more accessible and affordable way to fabricate PCBs. Using a similar fabrication process, we present our paper circuit design and fabrication technique based on the idea of subtractive processes.

## 3 PROPOSED TECHNIQUES

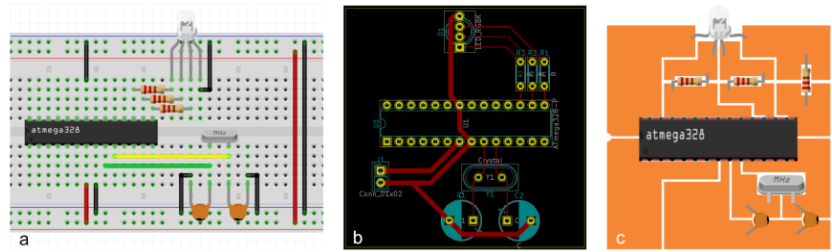

Figure 1: (a) Circuit diagram for an RGB LED in Fritzing, (b) traditional PCB layout for the circuit, and (c) circuit converted to paper circuit layout by PaperCAD

Figure 1 illustrates the process of converting a conventional circuit into a paper circuit design suitable for processing. Unlike conventional PCBs, this paper circuit features large conductive zones (orange area in (c)) rather than thin traces (red lines in (b)) connecting various components. This design enhances the conductive zone's area, thereby minimizing the amount of material that must be "subtracted" during the subtraction process. Although we had been doing this conversion manually in the beginning, we understood that the process was quite tedious. For this reason, we have developed PaperCAD software to help make this process easier for users. To begin the development of paper circuits using PaperCAD, we first need to conceptualize and create a basic circuit design with all the components. This can be done in Fritzing [Knörig, 2009], a commonly used tool for circuit design, which allows the user to export a netlist for the circuit in XML format. Our software tool converts this netlist into a circuit layout in vector format. For advanced users seeking greater customizability with their circuit designs, this technique also allows the user to develop the circuit layout design on a more advanced PCB design tool like KiCad or EagleCAD, and export a layer of the PCB as a vector file.

To fabricate the paper circuit, we can use the graphic files with a desktop vinyl cutter as a tool. First, we put a layer of conductive material, such as copper foil tape and fabric condition tape, on the paper. Then we import the design into the software for the vinyl cutter and cut it. Then we weed off the excess and get the final paper circuit. After getting the conductive layer of the paper circuit, we need to add the electronic components to it. In addition to the methods that have been used in previous work, such as soldering and using 3M conductive tape, the new method we propose is to staple the paper circuit and then let the pins of the electronic components go through. We call this method the stapler method, as those components are pinned to the paper like staples. Figure

2 shows the RGB LED unit we made using this technique. This technique, which does not require soldering, can further reduce the difficulty of making paper circuits.

In addition, we can also use fine tape to complete the fabrication. We first need to select "using fine tape" in PaperCAD, which will convert a paper circuit design that is more suitable for this fabrication method. Then we need to print the converted drawing, apply the fine tape to the blank areas, and cover them with conductive material. Due to the specificity of this method, the top layer of conductive material needs to be easily peelable, such as copper foil tape. Conductive fabric tape is not suitable for this fabrication method. Then we need to press on the top layer of material to make sure it is firmly adhered. Finally, we remove the fine tape in the middle to cut and weed the upper layer conductive material. For electronic components that have fewer pins, we use copper tape to attach them directly to the circuit. To make it easier to add copper tape, we also convert the tape rollers into copper tape dispensers. Figure 3 demonstrates the process we used to make the battery unit with this approach.

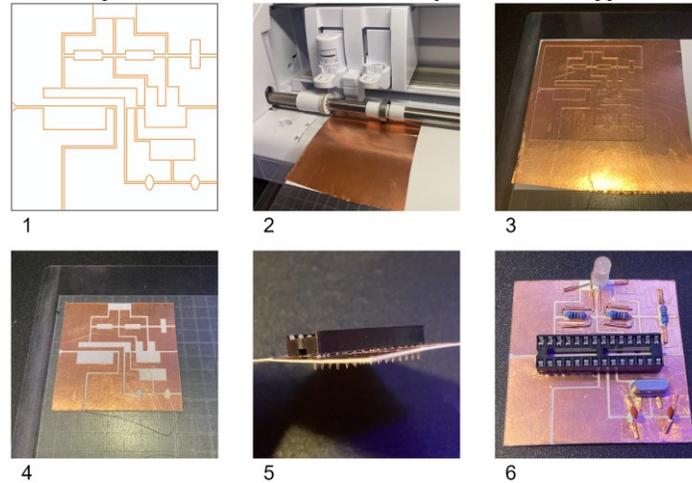

Figure 2: Demonstration of the fabrication process of a RGB LED unit (Step 1: Import the converted design into the vinyl cutter software. Step 2: Place the paper with copper foil attached to the cutting mat and load the vinyl cutter. Step 3: Make the cut. Step 4: Weed the excess parts. Step 5: Mount the IC socket using the stapler method. Step 6: Mount the other electronics using copper tape)

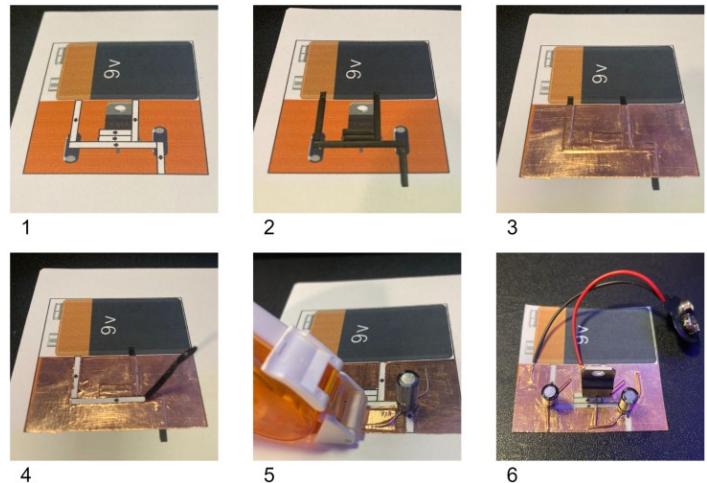

Figure 3: Demonstration of the fabrication process of a battery unit (Step 1: Print the converted design. Step 2: Apply the fine tape in the blank space. Step 3: Apply a layer of copper foil on the top. Step 4: Remove the fine tape, thus removing the copper foil on top of it. Step 5: Use the copper tape dispenser to mount the electronics. Step 6: Cut to the desired size)

## 4 DISCUSSION AND CONCLUSION

User needs and preferences play an important role in the development of electronic prototyping kits, one of the most critical needs is to rapidly iterate on a circuit. Our proposed design and fabrication techniques address this need by enabling users to quickly design, fabricate, and test circuits without waiting for PCB manufacturing or relying on hazardous chemicals. Our technology provides a safe, fast, and environmentally friendly electronic prototyping process.

We are committed to making electronics and hardware development accessible for more people, regardless of their age or skill level, which is the motivation behind our development of PaperCAD software and our proposed fabrication methods. The proposed techniques are designed to be easy to learn, without the need for specialized skills or equipment, and to provide better support for beginners in circuit design and fabrication. While we are still completing the development of PaperCAD, our expectation is that it will not only convert other circuit files, but also provide the option for users to generate circuit designs directly in PaperCAD using modular elements, further reducing the complexity of the process. Prior to this work, we proposed EdBoard [Yang, 2020], an educational breadboard design for children that offers a transparent plastic design with magnetic connections to introduce circuits to children. EdBoard is a successful building platform that has attracted a lot of children's interest in STEM. Building on that, we have reduced the cost of building by using paper, a much less expensive and ubiquitous material, while also making it more accessible for makers, educators, and others to fabricate the kit themselves using the described technique. We are developing PaperCAD as an Open Source project so that other researchers and educators can freely use and build upon it.

The future of electronic prototyping should be inclusive, allowing a wider range of people to be engaged, and we are excited to discuss and envision this future at the Beyond Prototyping Boards workshop. Our goal is to develop accessible and affordable electronic prototyping platforms with tools and techniques that can support people with different abilities, skill levels, and resources. Through this workshop, we aim to share our approach towards this goal with the community of designers, engineers, makers and researchers interested in these platforms, and learn from their innovative methods towards developing new paradigms of electronic toolkits.

**ACKNOWLEDGMENTS**

This material is based upon work supported by the U.S. National Science Foundation under Grant No IIS-2040489.

**REFERENCES**

Buechley, L., Hendrix, S., and Eisenberg, M. 2009. Paints, paper, and programs: first steps toward the computational sketchbook. In Proceedings of the 3rd International Conference on Tangible and Embedded Interaction (TEI '09), 9-12. https://doi.org/10.1145/1517664.1517670

Coombs Jr., C.F. Printed Circuits Handbook, Sixth Edition (The McGraw-Hill Companies, 2008). https://www.accessengineeringlibrary.com/content/book/9780071467346

Haider, G. 2013. Drawing Circuits. https://gottfriedhaider.com/Drawing-Circuits

Knörig, A., Wettach, R., and Cohen, J. 2009. Fritzing: a tool for advancing electronic prototyping for designers. In Proceedings of the 3rd International Conference on Tangible and Embedded Interaction (TEI '09), 351-358. https://doi.org/10.1145/1517664.1517735

Li, H., Brockmeyer, E., Carter, E.J., Fromm, J., Hudson, S.E., Patel, S.N., and Sample, A. 2016. PaperID: A Technique for Drawing Functional Battery-Free Wireless Interfaces on Paper. In Proceedings of the 2016 CHI Conference on Human Factors in Computing Systems (CHI '16), 5885-5896. https://doi.org/10.1145/2858036.2858249

Qi, J. and Buechley, L. 2010. Electronic popables: exploring paper-based computing through an interactive pop-up book. In Proceedings of the 4th International Conference on Tangible, Embedded, and Embodied Interaction (TEI '10), 121-128. https://doi.org/10.1145/1709886.1709909

Qi, J., Freed, N., Tseng, T., Shaw, F., Liedahl, B., Glowacki, B.R., and Kawahara, Y. 2021. Exquisite Circuits: Collaborative Electronics Design through Drawing Games. In Creativity and Cognition (C&C '21). Association for Computing Machinery, New York, NY, USA, Article 18, 1-14. https://doi.org/10.1145/3450741.3466776

Saul, G., Xu, C., and Gross, M.D. 2010. Interactive paper devices: end-user design & fabrication. In Proceedings of the 4th International Conference on Tangible, Embedded, and Embodied Interaction (TEI '10), 205-212. https://doi.org/10.1145/1709886.1709924

Yang, R., Candler, C., and Do, E.Y.-L. 2020. EdBoard: an educational breadboard. In Proceedings of the 2020 ACM Interaction Design and Children Conference: Extended Abstracts (IDC '20). Association for Computing Machinery, New York, NY, USA, 193-198. https://doi.org/10.1145/3397617.3397832